\documentclass[aps,pra,twocolumn,preprintnumbers,amsmath,amssymb,
superscriptaddress,floatfix,a4paper]{revtex4}%

\usepackage{graphicx}
\usepackage{dcolumn}
\usepackage{amsmath}
\usepackage{amssymb}
\usepackage{color}
\usepackage{multirow}
\usepackage{url}
\usepackage{bm}
\usepackage{xfrac}

\usepackage[utf8]{inputenc}
\usepackage{siunitx} 
\usepackage{verbatim} 
\usepackage[normalem]{ulem}

\usepackage{placeins} 
\usepackage{float} 

\usepackage{pdfpages}

\usepackage[hidelinks]{hyperref}
\hypersetup{
colorlinks=true,    
linkcolor=blue,
citecolor=blue,
urlcolor=blue,
pdfborder={0 0 0},
bookmarksopen=true,
bookmarksopenlevel=2,
pdfpagemode=UseOutlines,
pdfstartview={Fit},
pdftitle={Manuscript: Parallel spin-momentum locking in a chiral topological semimetal},
pdfauthor={J. A. Krieger},
pdfsubject={spin-PtGa},
pdfkeywords={PtGa,spin-ARPES,density-functional 
theory},
pdfhighlight= /I
}

\newcommand{\refsubfig}[2]{\hyperref[#1]{\ref*{#1}#2}}

\begin{document}

\title{Parallel spin-momentum locking in a chiral topological semimetal}

\author{Jonas~A.~Krieger}
\altaffiliation{These authors contributed equally to this work}
\affiliation{Max Planck 
Institut f\"ur 
Mikrostrukturphysik, 
Weinberg 2, 06120 Halle, Germany}

\author{Samuel~Stolz}
\altaffiliation{These authors contributed equally to this work}
\affiliation{Department of Physics,
University of California,
Berkeley, CA, USA}
\affiliation{nanotech@surfaces Laboratory, Empa,
Swiss Federal Laboratories for Materials Science and Technology,
8600 Dübendorf, Switzerland}

\author{Iñigo~Robredo}
\altaffiliation{These authors contributed equally to this work}
\affiliation{Max Planck Institute for Chemical Physics of Solids,
Dresden}
\affiliation{Donostia International Physics Center,
20018 Donostia - San Sebastian, Spain}

\author{Kaustuv~Manna}
\affiliation{Max Planck Institute for Chemical Physics of Solids,
Dresden
}
\affiliation{Indian Institute of Technology-Delhi, Hauz Khas, New Delhi 110 016, India}

\author{Emily~C.~McFarlane}
\affiliation{Max Planck 
Institut f\"ur 
Mikrostrukturphysik, 
Weinberg 2, 06120 Halle, Germany}

\author{Mihir~Date}
\affiliation{Max Planck 
Institut f\"ur 
Mikrostrukturphysik, 
Weinberg 2, 06120 Halle, Germany}

\author{Eduardo~B.~Guedes}
\affiliation{Swiss Light Source,
Paul Scherrer Institute,
5232 Villigen PSI, Switzerland}
\affiliation{Institut de Physique,
École Polytechnique Fédérale de Lausanne,
1015 Lausanne, Switzerland}

\author{J.~Hugo~Dil}
\affiliation{Swiss Light Source,
Paul Scherrer Institute,
5232 Villigen PSI, Switzerland}
\affiliation{Institut de Physique,
École Polytechnique Fédérale de Lausanne,
1015 Lausanne, Switzerland}

\author{Chandra~Shekhar}
\affiliation{Max Planck Institute for Chemical Physics of Solids,
Dresden
}

\author{Horst~Borrmann}
\affiliation{Max Planck Institute for Chemical Physics of Solids,
Dresden
}
\author{Qun~Yang}
\affiliation{Max Planck Institute for Chemical Physics of Solids,
Dresden
}
\author{Mao~Lin}
\affiliation{Department of Physics,
University of Illinois,
Urbana-Champaign, USA}

\author{Vladimir~N.~Strocov}
\affiliation{Swiss Light Source,
Paul Scherrer Institute,
5232 Villigen PSI, Switzerland}

\author{Marco~Caputo}
\affiliation{Swiss Light Source,
Paul Scherrer Institute,
5232 Villigen PSI, Switzerland}

\author{Banabir~Pal}
\affiliation{Max Planck Institut f\"ur 
Mikrostrukturphysik, Weinberg 2, 06120 Halle, Germany}

\author{Matthew~D.~Watson}
\affiliation{Diamond Light Source,
Didcot OX11 0DE, UK}

\author{Timur~K.~Kim}
\affiliation{Diamond Light Source,
Didcot OX11 0DE, UK}

\author{Cephise~Cacho}
\affiliation{Diamond Light Source,
Didcot OX11 0DE, UK}

\author{Federico~Mazzola}
\affiliation{CNR-IOM, Area Science Park, Strada Statale 14 km 163.5, I-34149 Trieste, Italy}

\author{Jun~Fujii}
\affiliation{CNR-IOM, Area Science Park, Strada Statale 14 km 163.5, I-34149 Trieste, Italy}

\author{Ivana~Vobornik}
\affiliation{CNR-IOM, Area Science Park, Strada Statale 14 km 163.5, I-34149 Trieste, Italy}

\author{Stuart~S.P.~Parkin}
\affiliation{Max Planck Institut f\"ur 
Mikrostrukturphysik, Weinberg 2, 06120 Halle, Germany}

\author{Barry~Bradlyn}
\affiliation{Department of Physics,
University of Illinois,
Urbana-Champaign, USA}

\author{Claudia~Felser}
\affiliation{Max Planck Institute for Chemical Physics of Solids,
Dresden
}
\author{Maia~G.~Vergniory}
\affiliation{Max Planck Institute for Chemical Physics of Solids,
Dresden
}
\affiliation{Donostia International Physics Center,
20018 Donostia - San Sebastian, Spain}

\author{Niels~B.~M.~Schr\"oter}
\email{niels.schroeter@mpi-halle.mpg.de}
\affiliation{Max Planck Institut f\"ur 
Mikrostrukturphysik, Weinberg 2, 06120 Halle, Germany}

\begin{abstract}
Spin-momentum locking in solids describes a directional relationship between the electron’s spin angular momentum and its linear momentum over the entire Fermi surface. While orthogonal spin-momentum locking, such as Rashba spin-orbit coupling~\cite{bychkov_properties_1984}, has been studied for decades and inspired a vast number of applications~\cite{manchon_new_2015}, its natural counterpart, the purely parallel spin-momentum locking, has remained elusive in experiments. Recently, chiral topological semimetals that host single- and multifold band crossings have been predicted to realize such parallel locking~\cite{chang_topological_2018,lin_spin-momentum_2022}. Here, we use spin- and angle-resolved photoelectron spectroscopy to probe spin-momentum locking of a multifold fermion in the chiral topological semimetal PtGa via the spin-texture of its topological Fermi-arc surface states. We find that the electron spin of the Fermi-arcs points orthogonal to their Fermi surface contour for momenta close to the projection of the bulk multifold fermion, which is consistent with parallel spin-momentum locking of the latter. We anticipate that our discovery of parallel spin-momentum locking of multifold fermions will lead to the integration of chiral topological semimetals in novel spintronic devices, and the search for spin-dependent superconducting and magnetic instabilities in these materials.
\end{abstract}

\maketitle
\noindent 
Objects that lack inversion and mirror symmetries are chiral and often show an intimate relationship between the direction of their spin angular momentum and their linear momentum. For instance, in particle physics, the chirality of the hypothetical massless Weyl-fermion determines if its spin is either parallel or antiparallel to its momentum direction~\cite{armitage_weyl_2018}. In chemistry and biology, chiral molecules such as DNA have been shown to act as effective spin polarisers, via an effect known as chirality-induced-spin-selectivity (CISS)~\cite{naaman_spintronics_2015,Liang2022}. CISS has recently been attributed to a locking of the electron momentum to the electron's orbital angular momentum, which is linked to the spin via spin-orbit coupling (SOC)~\cite{liu_chirality-driven_2021}. In condensed matter physics, crystalline symmetries lock the spin-direction of electrons along the direction of screw- and rotational- axes, and orthogonal to mirror planes \cite{hirayama_weyl_2015}. It has been argued that in structurally chiral crystals such as trigonal tellurium, the Fermi surface would generically show parallel spin-momentum locking (SML) due to the absence of mirror planes. However, this is strictly true only for momentum directions along such screw- or rotational-axes, but not along arbitrary momentum directions, and substantial deviations from parallel SML can be observed in ab-initio calculations around the valence band maximum in trigonal tellurium~\cite{sakano_radial_2020,gatti_radial_2020}. Finding experimental evidence for truly parallel SML is of fundamental interest because it is the natural counterpart to Rashba type orthogonal SML (Fig.~1a) that has been studied for decades and has inspired a vast number of fundamental discoveries and practical applications~\cite{manchon_new_2015}. Similarly, parallel SML would enable many novel phenomena and applications, such as longitudinal magnetoelectric effects and spin-orbit torques for switching of magnetic layers with perpendicular magnetic anisotropy \cite{he_kramers_2020} or new types of correlated phenomena, such as exotic superconductivity \cite{link_d-wave_2020,boettcher_interplay_2020,fu2008superconducting,bornemann2019complex,frigeri2004superconductivity}.

\begin{figure*}[hbt]
  \includegraphics[width=0.95\linewidth]{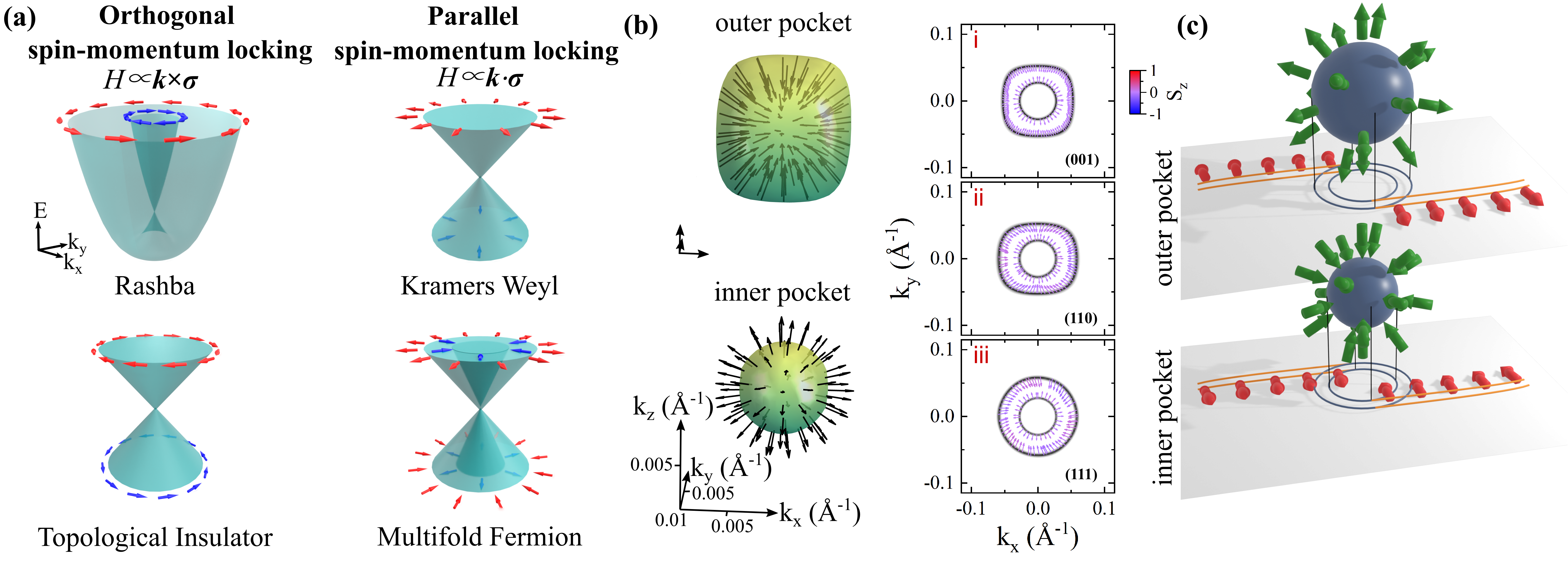}
  \caption{\textbf{Spin-momentum locking prototypes and spin-texture of multifold fermion in PtGa.} 
  (a) Overview of the two prototypical forms of spin-momentum locking (SML) (left) Rashba type orthogonal SML locking and (right) Kramers-Weyl type parallel SML locking.
  (b) Calculated three-dimensional spin-texture of the two spherical Fermi surfaces formed by the S~=~$\sfrac{3}{2}$ Rarita-Schwinger multifold fermion in PtGa. Panels (i-iii) show two-dimensional slices through the three-dimensional spin-texture, indicating the radial nature of the spin-texture along all momentum directions.
  (c) Illustration of the link between the spin-texture of the bulk multifold fermions (green arrows) and the spin-texture of the topological Fermi-arcs (red arrows). Since the Fermi-arcs are tangential to the projected multifold Fermi surface pockets and inherit their radial spin-texture, the spin-texture in the arc is orthogonal to its Fermi surface contour.}
\label{fig:fig1}
\end{figure*}

Parallel SML has been predicted in chiral topological semimetals, which are a recently discovered class of quantum materials that host singly or multiply degenerate electronic nodal band crossings at time-reversal invariant momenta (TRIM) of the Brillouin zone close to the Fermi level with large topological charges \cite{bradlyn_beyond_2016,takane_observation_2019,rao_observation_2019,sanchez_topological_2019,schroter_chiral_2019, schroter_observation_2020}. The effective Hamiltonian describing such singly degenerate crossings (known as Kramers-Weyl fermions) takes to leading order in a $k \cdot p$ expansion the form of $H \propto \bm{k} \cdot \bm{\sigma}$, where $\bm{k}$ is the crystal momentum wave vector and $\bm{\sigma}$ is a vector of Pauli matrices acting on the true electron spin. Such quasiparticles would therefore provide perfect parallel SML along all momentum directions on the Fermi surface as long as the node is close enough to the Fermi level such that the $k \cdot p$ approximation holds. Note that in contrast to Kramers-Weyl fermions, Weyl-fermions away from TRIM, which have been found in many achiral materials, are typically described by Pauli matrices that act on orbital pseudospin degrees of freedom, and thus do not exhibit spin-momentum locking \cite{chang_topological_2018}. Recently, we developed a theory that shows that Fermi surfaces formed by multifold fermions also possess spin-momentum locking close to the degeneracy point ~\cite{lin_spin-momentum_2022}. Whilst various complex spin-textures are possible depending on the involved orbitals, the simplest possible spin-texture shows, as in Kramers-Weyl fermions, perfect parallel SML over the entire Fermi surface. Additionally, Refs.~\cite{mera_acosta_different_2021,Tan_unified_2022} considered the spin texture in chiral crystals for bands that can be treated in the limit of weak SOC, and derived the possible spin textures for weakly spin-split Fermi surfaces, finding many cases where chiral cubic symmetry enforces perfectly parallel SML along generic momentum directions. In the context of multifold fermions, Refs.~\cite{mera_acosta_different_2021,Tan_unified_2022} apply to Fermi pockets far enough from the nodal point such that bands are only weakly spin-split; this complements the approach of Ref.~\cite{lin_spin-momentum_2022} which is applicable much closer to the nodal point, where bands cannot be described in a simple spin-orbit decoupled basis.

Multifold fermions that could realize parallel SML are the fourfold degenerate spin S~=~$\sfrac{3}{2}$ Rarita-Schwinger mutifold fermion and the sixfold degenerate S=1 multifold fermion, which were predicted to occur at the $\Gamma$ and $R$ point, respectively, in materials that belong to the chiral cubic B20 crystal structure (space group 198)~\cite{bradlyn_beyond_2016,chang2017unconventional,flicker2018chiral,Tang2017}. Recent experiments confirmed the existence of the multifold fermions in several B20 materials, such as CoSi~\cite{takane_observation_2019,rao_observation_2019}, RhSi~\cite{sanchez_topological_2019}, PdGa~\cite{schroter_observation_2020}, AlPt~\cite{schroter_chiral_2019}, or PtGa~\cite{yao_observation_2020,ma_observation_2021} by  directly imaging the nodal crossings and determining the Chern number by counting the Fermi-arc surface states that connect the surface projections of multifold fermions at $R$ and $\Gamma$ using angle-resolved photoelectron spectroscopy (ARPES). Moreover, a direct link between the sign of the Chern numbers and the handedness of the host crystal has been established \cite{schroter_observation_2020,sessi_handedness-dependent_2020}.

The method of choice to study momentum resolved spin-textures is spin-resolved ARPES (spin-ARPES). However, spin-ARPES measurements of Kramers-Weyl and multifold fermions are challenging for two main reasons: Firstly, accurate measurements of bulk band structures around 3D nodal points typically requires using soft X-ray ARPES to obtain a high out-of-plane momentum resolution. Because of the low photoemission cross-section in the soft X-ray energy range, spin-resolved soft X-ray ARPES with the currently available single-channel spin-detectors is typically unfeasible due to low count rates. Secondly, in many materials, the separation between Kramers-Weyl or multifold fermions and other spin-polarized bands is small in energy and momentum, which makes it difficult to disentangle the contributions of all individual bands due to the limited resolution of currently available spin-detectors. For instance, for the material PtGa, previous studies reported the spin-polarization of the bulk states consistent with time-reversal symmetry ~\cite{ma_observation_2021}, but no signatures of spin-momentum locking.

In the present work, we overcome these obstacles by exploiting the close relationship between the multifold fermion bulk states and their topological Fermi-arc surface states in the chiral topological semimetal PtGa by combining ab-initio calculations and spin-resolved and spin-integrated ARPES. Our ab-initio calculations predict that for PtGa the multifold fermion at the $\Gamma$ point is located just below the Fermi level and forms two electron pockets on the Fermi surface (Fig.~1b): the inner one is practically spherical and shows parallel SML locking along all momentum directions. The outer one is almost spherical, with small corrections towards a rectangular shape that lead to minor deviations from parallel SML. Since the bands making up the multifold fermion at $\Gamma$ are only weakly split by SOC (see supplementary Fig.~S5 for a band structure calculation without SOC), parallel SML is also consistent with the prediction of Refs.~\cite{mera_acosta_different_2021,Tan_unified_2022} for B20 crystal structure.

Turning to the surface states, we note that the Fermi-arcs connect and simultaneously are derived from the surface-projected Fermi surface pockets of multifold fermions with opposite Chern number \cite{okugawa_dispersion_2014}. In the absence of spin-orbit coupling, there will be two spin-degenerate Fermi-arcs emanating from the bulk projection of the Fermi surface. Close to the surface projection of the multifold point, we expect the Fermi-arcs to be split by bulk SOC, since for this region of momentum space the wave functions for the Fermi-arcs extend deep into the bulk. If the bulk SOC favors parallel SML, we therefore expect that close to the projected Fermi surface of the multifold fermions the Fermi-arcs, which disperse tangentially to the bulk projected pockets \cite{okugawa_dispersion_2014}, will have their spin oriented perpendicular to their contour (illustrated in Fig.~1c). By measuring the surface spin-texture of the Fermi-arcs sufficiently close in momentum to the projection of the multifold node, we can therefore infer the spin-texture of the latter. For larger momenta away from the multifold fermion, the $k \cdot p$ approximation will break down and the Fermi-arc spin texture will no longer be linked to the spin-texture of the multifold fermions. Because the Fermi-arcs are non-dispersive in the out-of-plane direction, spin-ARPES can be measured at lower photon energies with higher cross-section, which makes such experiments feasible with currently available synchrotron set-ups. Furthermore, since there are sufficiently large projected bulk band gaps at the Fermi surface of PtGa close to the multifold fermion at the $\Gamma$ point, we can measure the Fermi-arc spin-texture without interference from other bands.

The samples studied in our work were enantiopure PtGa single-crystals that were cut along different crystallographic directions, then polished and cleaned in vacuum by cycles of Ar-ion sputtering and annealing (see Methods section for information about sample characterization, preparation, and measurement conditions). Our bulk sensitive soft X-ray ARPES measurements on PtGa crystals cut along the [111]-direction directly image the multifold band crossing at the $R$ point (Fig.~2a-b) and at the $\Gamma$ point (Fig.~2c-d), which are well reproduced by our ab-initio calculations. Note that the Rarita-Schwinger fermion at the $\Gamma$ point is only about $\SI{100}{meV}$ below the Fermi level, which means that the Fermi wave vector $k_F$ is close to the $\Gamma$ point and thus the SML derived from the $k \cdot p$ approximation can be expected to hold on the Fermi surface.

\begin{figure*}[hbt]
  \includegraphics[width=0.99\linewidth]{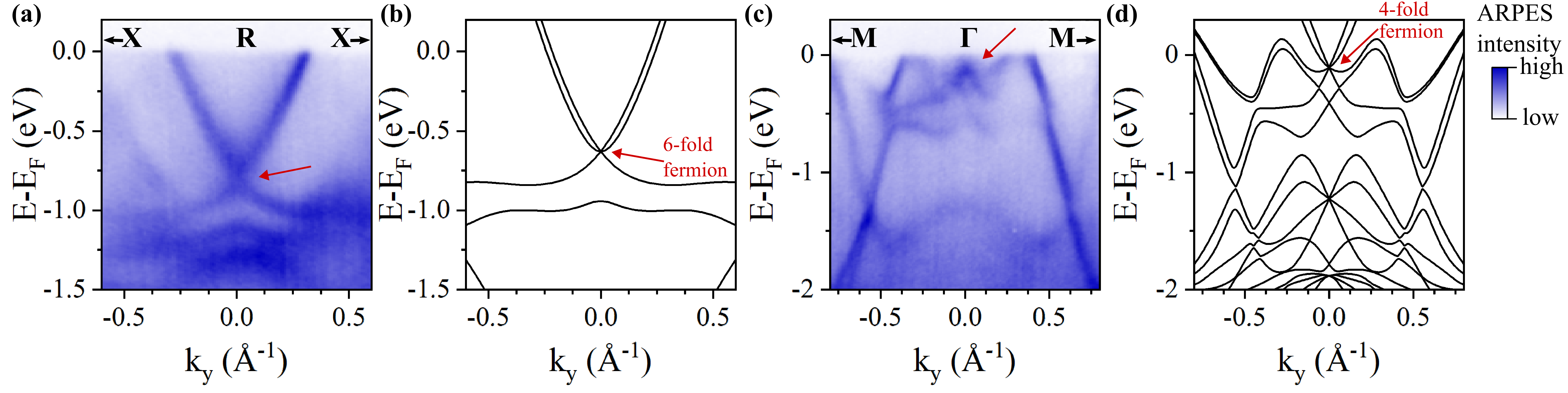}
  \caption{\textbf{Spin-integrated bulk band structure of PtGa(111).} 
  (a) Soft X-ray ARPES  band dispersion along the X\,-\,R\,-\,X direction measured with $h\nu=\SI{547}{eV}$ and $p$-polarization. The red arrow indicates the 6-fold band crossing located at the R point, which is well-reproduced by ab-initio calculations displayed in (b).
  (c) Soft X-ray ARPES band dispersion along the M\,-\,G\,-\,M direction measured with $h\nu=\SI{653}{eV}$ and $p$-polarization. The red arrow indicates the 4-fold Rarita-Schwinger multifold fermion located at the G point that is located about 100~meV below the Fermi level, which is well reproduced by our ab-initio calculations (d). }
\label{fig:fig2}
\end{figure*}
With our more surface sensitive VUV-ARPES measurements of the (001) surface of PtGa (Fig.~3) we resolve two Fermi-arcs close to the center of the surface Brillouin zone that have a Fermi surface contour approximately parallel to the $k_x$ axis, and which merge into the small projected bulk pockets of the multifold fermion at the $\Gamma$ point (upper red arrow in Fig.~3a). Further away from the $\Gamma$ point, these Fermi-arcs wind from the center of the surface Brillouin zone around the projected bulk pocket centered at the $R$ point and ultimately merge with the latter. The Fermi-arcs are well reproduced by our ab-initio calculations (Fig.~3b) and our photon energy dependent VUV-ARPES measurement confirms experimentally that they have no out-of-plane dispersion (Fig.~3c), as expected for 2D surface states. Experimental band dispersions along high-symmetry directions are displayed in Fig.~3d-e, which show how four Fermi-arcs merge with the projection of the multifold fermion at $\overline{\Gamma}$ point in the center of the surface Brillouin zone.

\begin{figure*}[hbt]
  \includegraphics[width=0.95\linewidth]{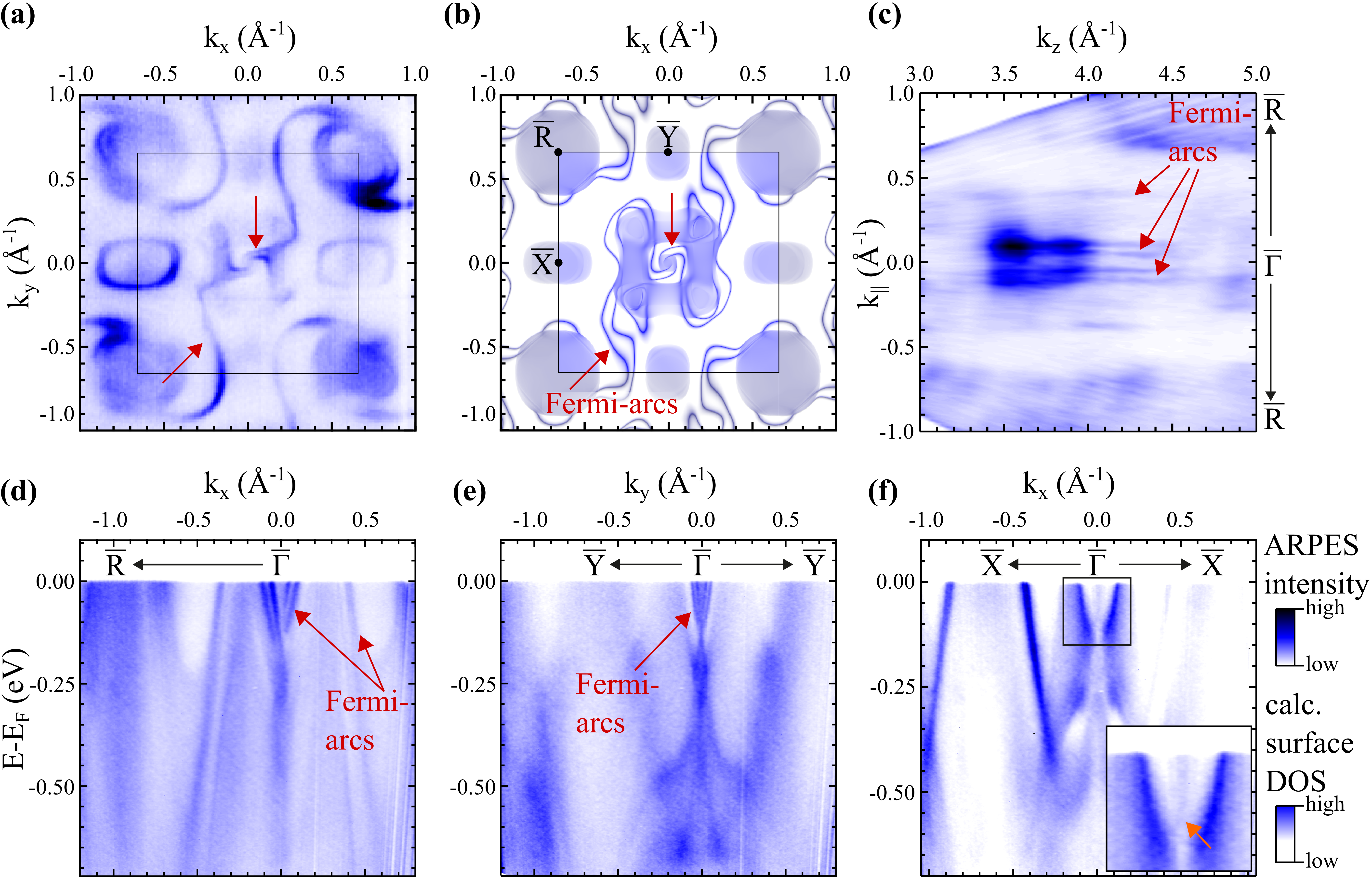}
  \caption{\textbf{Spin-integrated Fermi-arc surface state dispersion on the (001) surface of PtGa.}
  (a) Fermi surface measured with $h\nu=\SI{60}{eV}$ and $p$-polarization. Red arrows indicate the Fermi-arcs close to the center and at the boundary of the Brillouin zone.
  (b) Ab-initio calculated Fermi surface of PtGa. 
  (c) Fermi surface along the in-plane $k_{\parallel}$ vs. out-of-plane $k_{z}$ momentum direction.
  (d-f) Band dispersions along various high symmetry directions, measured with $h\nu=\SI{60}{eV}$ and $p$-polarization. Inset in (f) shows multifold fermion at $\overline{\Gamma}$ indicated by orange arrow.}
\label{fig:fig3}
\end{figure*}

The resolution of the four individual Fermi-arcs close to the multifold fermion at $\overline{\Gamma}$, which are well separated from other bands due to a large projected bulk band gap, enables us to measure their spin-texture (Fig.~4). As illustrated in Fig.~4a, we find that close to $\overline{\Gamma}$, the spin polarization of the outer Fermi-arc points anti-parallel to the $k_y$ axis, whilst the inner Fermi-arc points parallel to the $k_y$ axis. For both Fermi-arcs, the spins point orthogonal to the Fermi contour, which is also well reproduced by our ab initio calculations (Fig.~4b). Since the Fermi-arcs inherit the spin-texture of the multifold fermion for momenta close to $\overline{\Gamma}$, we can infer from the spin-texture of the Fermi-arcs that the inner (outer) bulk pocket of the multifold fermion has a spin texture that is parallel (antiparallel) to the momentum pointing away from (towards) the $\Gamma$ point. This is consistent with parallel spin-momentum locking of the multifold fermion predicted by our bulk ab-initio calculations.

\begin{figure*}[hbt]
  \includegraphics[width=0.99\linewidth]{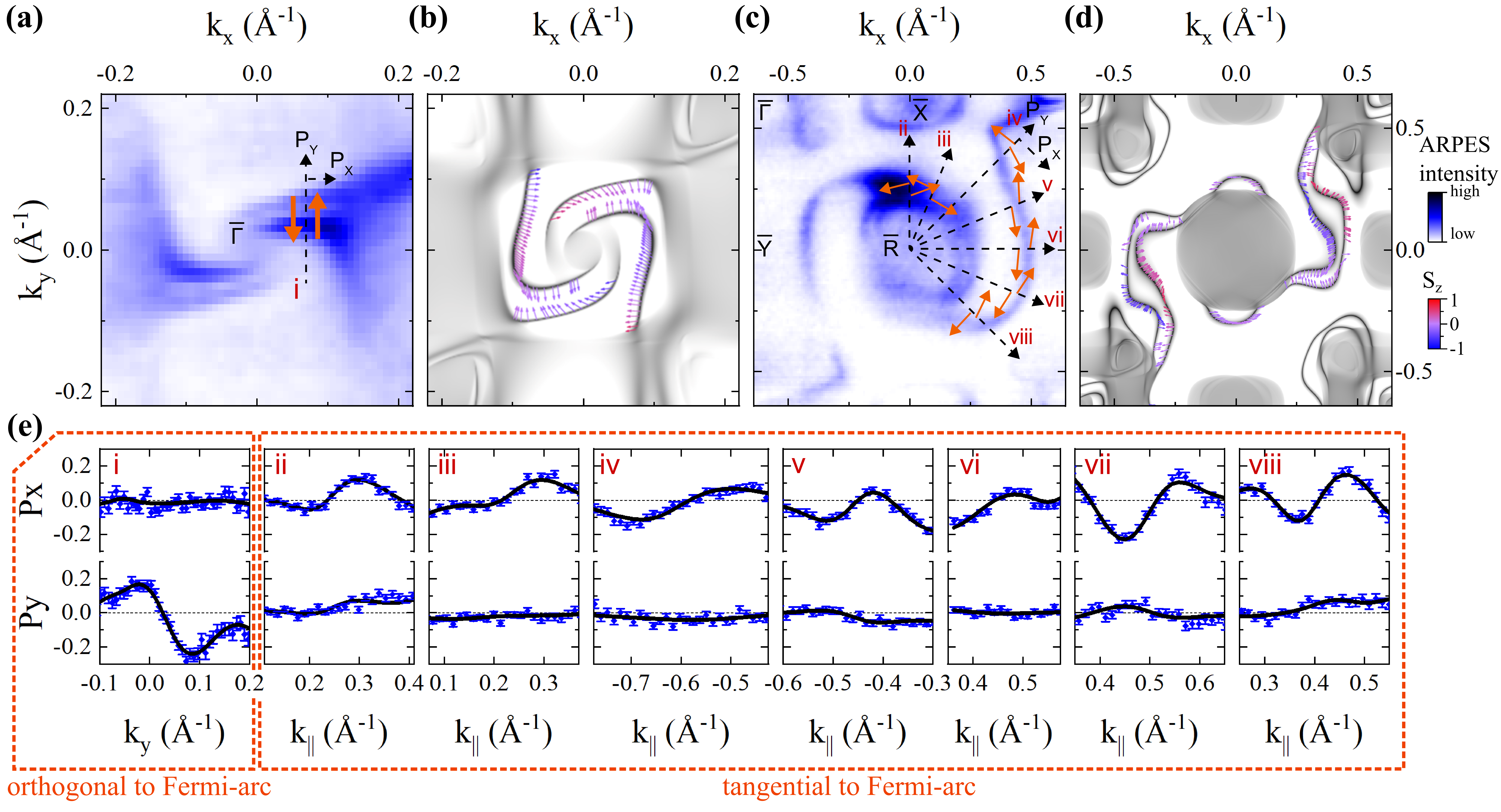}
  
  \caption{\textbf{Spin-texture of the Fermi-arcs on the (001) surface of PtGa.}
   (a) Spin-texture of the two Fermi-arcs close to the projection of the multifold fermion at the $\overline{\Gamma}$ point, orange arrows indicate the direction of the in-plane spin-polarization.
  (b) Calculation of the spin-texture of the Fermi-arcs around the $\overline{\Gamma}$ point, in good agreement with the experimental result.
  (c)  Spin-texture of the two Fermi-arcs dispersing in-between $\overline{\Gamma}$ and $\overline{R}$ points. The orange arrows indicate the direction of the in-plane spin-polarization.
  (d) Calculation of the Fermi-arc spin-texture in-between $\overline{\Gamma}$ and $\overline{R}$.
  (e) Asymmetries in the measured in-plane spin-polarizations along various momentum directions indicated by the Roman numerals in (a,c).}
\label{fig:fig4}
\end{figure*}

We have also measured the spin-texture of the Fermi-arc in-between  $\overline{\Gamma}$ and $\overline{R}$ (shown in Fig.~4c-e), expecting that there would be no spin-momentum locking due to a breakdown of the $k \cdot p$ approximation. Surprisingly, we find that the measured in-plane spin-texture is predominantly tangential to the Fermi contour. To explain why the Fermi-arc spin-texture appears in such a fine-tuned state far from high symmetry points, further theoretical work is required.

To check that the measured spin-texture is indeed a property of the initial states and not an artifact from matrix element effects in the photoemission process, we have repeated several measurements for different photon energies and light polarizations. We always obtained the same result for the direction of the spin-texture (see supplementary Fig.~S3), which verifies that the spin-texture is a property of the initial state. As another consistency check, we have also investigated the spin-texture of the Fermi-arc of the opposite enantiomer and found that the spin-texture behaves as expected from the transformational properties of a pseudovector (see supplementry Fig.~S4).

In conclusion, our combined spin-integrated and spin-resolved ARPES measurements in combination with our ab-initio calculations provide strong evidence that the Rarita-Schwinger multifold fermion at the $\Gamma$ point in PtGa forms a Fermi surface with parallel spin-momentum locking. This finding makes chiral topological semimetals that host multifold fermions interesting candidates to realize  a large longitudinal magnetoelectric effect (LME) that could be used for novel spintronic devices \cite{he_kramers_2020}. The perfectly radial spin-momentum locking ensures that the LME would be isotropic, which suggests that even rotationally disordered films would be suitable for applications because the direction of the resulting spin-orbit torque would only depend on the direction of the applied field and be independent of the crystallographic direction of individual grains. Moreover, a Fermi surface with parallel SML can be expected to realize unconventional types of superconductivity, either by interfacing non-superconducting~B20 materials like PtGa with superconductors that can induce superconductivity via the proximity effect, or by looking at intrinsic superconductors that host multifold or Kramers-Weyl fermions \cite{gao_topological_2022}. Besides superconductivity, other Fermi surface instabilities could be affected by the perfectly radial spin-texture of multifold and Kramers-Weyl fermions, such as the charge density wave order that has been found in Kramers-Weyl candidates, such as $(TaSe_4)_2I$ \cite{kim_kramers-weyl_2021}.

\FloatBarrier
\bibliography{Biblio}

\clearpage

\section*{Acknowledgment}
 We acknowledge Diamond Light Source for time on beamline I05 under Proposals No.~SI20617-1, and No.~SI24703-1, Swiss Light source for beamtime on ADRESS and COPHEE under Proposals No. 20212139, 20210765, 20210484, 20191797 and Elettra Sincrotrone Trieste on APE-LE under Proposal No.~20215697. 
 J.A.K. acknowledges support by the Swiss National Science Foundation (SNF-Grant
No.~P500PT\_203159 ). 
S.S. acknowledges funding from the Swiss National Science Foundation under the project numbers 195133 and 159690. The authors thank Sulamith Gutwein and Theresa Voigt for help with the 3d plots in Fig.~1(a), and Wenxin~Li,  Procopi~C.~Constantinou, and Tianlun Yu for help during the ARPES measurements. 
B.B. acknowledges the support of the U.~S. National Science Foundation under grant DMR-1945058. M.G.V. and C.F. thanks support to the Deutsche Forschungsgemeinschaft (DFG, German Research Foundation) - FOR 5249 (QUAST). M.G.V. and I.R. acknowledge the Spanish Ministerio de Ciencia
e Innovacion (grant PID2019-109905GB-C21). 
K.M., C.S.~and C.F.~acknowledge financial support by European Research Council (ERC) Advanced Grant No.~742068 (“TOPMAT”), Deutsche Forschungsgemeinschaft (DFG) under SFB 1143 (Project No.~247310070) and Würzburg-Dresden Cluster of Excellence on Complexity and Topology in Quantum Matter-ct.qmat (EXC 2147, project no.~39085490). K.M.~acknowledges Max Plank Society for the funding support under Max Plank–India partner group project and Board of Research in Nuclear Sciences (BRNS) under 58/20/03/2021-BRNS/37084/ DAE-YSRA.
This work has been partly performed in the framework of the nanoscience foundry and fine analysis (NFFA-MUR Italy Progetti Internazionali) facility.

\section*{Competing Interest}
The authors declare no competing interest.

\section*{Data availability}
The data of this study will be made available on the Open Research Data Repository of the Max Planck Society. (reserved DOI: \href{https://dx.doi.org/10.17617/3.GIT9JN}{10.17617/3.GIT9JN} )

\section*{Methods}

\subsection*{Crystal growth}
PtGa single crystals were grown from the melt with the self-flux technique as described in Ref.~\cite{yao_observation_2020}. Polished surfaces along different high symmetry directions were produced, after aligning the crystals at room temperature with a white-beam backscattering Laue X-ray setup.

\subsection*{ARPES}

Prior to any ARPES experiments, the PtGa single-crystals were cleaned in ultrahigh vacuum with a base-pressure better than 5$\times$10$^{-9}$~mbar by repeated cycles of sputtering (Ar$^+$, 1 keV, 1$\times$10$^{-5}$~mbar) and annealing (970~K) until a clear LEED pattern was obtained.

Soft X-ray (SX-ARPES) experiments were performed at the SX-ARPES endstation~\cite{strocov_soft-x-ray_2014} of the ADRESS beamline~\cite{strocov_high-resolution_2010} at the Swiss Light Source, Switzerland. This endstation is equipped with a SPECS analyzer with an angular resolution of 0.07°. The photon energy was tuned between 350~eV and 650~eV and the combined energy resolution was ranging between 50~meV to 100~meV.

Spin-integrated vacuum ultraviolet ARPES (VUV-ARPES) measurements were executed at the high-resolution ARPES branch line of the beamline I05~\cite{hoesch_facility_2017} at the Diamond Light Source, United Kingdom. The endstation is equipped with a Scienta R4000 analyzer. Note that in the k$_z$ scan in Fig.~3(c) at $h\nu$ above \SI{50}{eV}, a linear drift of the Gamma point by less than \SI{1}{\degree}  was corrected manually by assuming the spectral features to be symmetric around Gamma, which they should be due to time-reversal symmetry.

Spin-resolved VUV-ARPES measurements close to the $\Gamma$-point (Fig.~4(d)i)  were carried out at APE-LE endstation at Elettra Synchrotrone Trieste, Italy, using a VLEED spin-detector~\cite{bigi_very_2017}. The MDCs were acquired with alternating magnetization of the spin-filter crystals in a $(+B,-B,-B,+B)$ sequence, to average out first-order systematics on the resulting asymmetry spectra. 

Spin-resolved ARPES measurements around the R-point (Fig. 4(d)ii-vii) were carried out at the COPHEE branch~\cite{hoesch_spin-polarized_2002} at the SIS endstation of the Swiss Light Source, Switzerland, which is equipped with double classical Mott detectors and has an angle and energy resolution better than \SI{1.5}{\degree} and \SI{75}{meV}, respectively.

All ARPES experiments were executed with the sample held around \SI{20}{K} and at a pressure better than \SI{2e-10}{mbar}. All the spin-resolved MDCs were measured perpendicular to the scattering plane.

For the calculation of the spin polarization, the instrumental asymmetry was corrected by numerically adjusting the detector relative gain between opposite Mott detectors/for opposite spin-filter magnetization. This will affect the absolute value of the polarization, but not its angle dependence.
To quantify the orientation of the Fermi-arc spin-polarization, we used a 3D vectorial fit of the measured spin expectation values with a well-established fitting routine~\cite{meier_measuring_2009}. For all Fermi-arcs, the magnitude of the spin expectation values were assumed to be 1. The Sherman factor has been set 0.3 and 0.08 for the data recorded at the APE-LE and COPHEE endstation, respectively. When transferring the spin polarization onto the Fermi surface, MDCs measured at the opposite side of the high symmetry point (indicated by negative $k_\parallel$-values) were moved to the other side by assuming time-reversal symmetry.

Note that in our measurements of the spin-texture, we cannot directly measure the out-of-plane spin component since the signal will be averaged over multiple domains with Fermi-arcs spin-polarized in opposite out-of-plane directions. The measured out-of-plane spin-texture is therefore always close to zero (see supplementary Figs.~S1,~S2 and supplement for more information).

    \subsection*{Ab-initio calculations}
In order to study the spin-polarization properties of the bulk electronic bands we performed density functional theory (DFT) calculations as implemented in the Vienna ab initio simulation package (VASP)~\cite{VASP1, VASP2, VASP3, VASP4}. The interaction between ion cores and valence electrons are treated by the projector augmented wave method~\cite{PAW}. We use the Perdew-Burke-Ernzerhof (PBE) parametrization~\cite{PBE} within the generalized gradient approximation (GGA) and spin-orbit coupling is taken into account by the second variation method~\cite{DFT-SOC}. We used a Monkhorst-Pack grid of (9$\times$9$\times$9) k-points for reciprocal space integration and 500eV as cut-off energy. Spin polarization data was extracted using the PyProcar python package \cite{PyProcar}.

To analyze the surface spectrum and Fermi arcs we construct an interpolated Hamiltonian in the maximally localized Wannier basis~\cite{Wannier90}. Since we are mainly interested in the spin-polarization, we construct maximally localized Wannier functions from a DFT calculation as implemented in the full-potential local-orbital code (FPLO)~\cite{FPLO}, which preserves the spin-polarization of the Wannier basis. The exchange correlation functional is treated within the (GGA) approximation. We then construct the Wannier Hamiltonian choosing as starting projections $s$, $p$ and $d$ orbitals for Pt and $s$, $p$ orbitals for Ga. The surface spectrum is computed following the iterative Green's function method as implemented in WannierTools~\cite{WannierTools}.

\end{document}


\author{Jonas~A.~Krieger}
\altaffiliation{These authors contributed equally to this work}
\affiliation{Max Planck 
Institut f\"ur 
Mikrostrukturphysik, 
Weinberg 2, 06120 Halle, Germany}

\author{Samuel~Stolz}
\altaffiliation{These authors contributed equally to this work}
\affiliation{Department of Physics,
University of California,
Berkeley, CA, USA}
\affiliation{nanotech@surfaces Laboratory, Empa,
Swiss Federal Laboratories for Materials Science and Technology,
8600 Dübendorf, Switzerland}

\author{Iñigo~Robredo}
\altaffiliation{These authors contributed equally to this work}
\affiliation{Max Planck Institute for Chemical Physics of Solids,
Dresden}
\affiliation{Donostia International Physics Center,
20018 Donostia - San Sebastian, Spain}

\author{Kaustuv~Manna}
\affiliation{Max Planck Institute for Chemical Physics of Solids,
Dresden
}
\affiliation{Indian Institute of Technology-Delhi, Hauz Khas, New Delhi 110 016, India}

\author{Emily~C.~McFarlane}
\affiliation{Max Planck 
Institut f\"ur 
Mikrostrukturphysik, 
Weinberg 2, 06120 Halle, Germany}

\author{Mihir~Date}
\affiliation{Max Planck 
Institut f\"ur 
Mikrostrukturphysik, 
Weinberg 2, 06120 Halle, Germany}

\author{Eduardo~B.~Guedes}
\affiliation{Swiss Light Source,
Paul Scherrer Institute,
5232 Villigen PSI, Switzerland}
\affiliation{Institut de Physique,
École Polytechnique Fédérale de Lausanne,
1015 Lausanne, Switzerland}

\author{J.~Hugo~Dil}
\affiliation{Swiss Light Source,
Paul Scherrer Institute,
5232 Villigen PSI, Switzerland}
\affiliation{Institut de Physique,
École Polytechnique Fédérale de Lausanne,
1015 Lausanne, Switzerland}

\author{Chandra~Shekhar}
\affiliation{Max Planck Institute for Chemical Physics of Solids,
Dresden
}

\author{Horst~Borrmann}
\affiliation{Max Planck Institute for Chemical Physics of Solids,
Dresden
}
\author{Qun~Yang}
\affiliation{Max Planck Institute for Chemical Physics of Solids,
Dresden
}
\author{Mao~Lin}
\affiliation{Department of Physics,
University of Illinois,
Urbana-Champaign, USA}

\author{Vladimir~N.~Strocov}
\affiliation{Swiss Light Source,
Paul Scherrer Institute,
5232 Villigen PSI, Switzerland}

\author{Marco~Caputo}
\affiliation{Swiss Light Source,
Paul Scherrer Institute,
5232 Villigen PSI, Switzerland}

\author{Banabir~Pal}
\affiliation{Max Planck Institut f\"ur 
Mikrostrukturphysik, Weinberg 2, 06120 Halle, Germany}

\author{Matthew~D.~Watson}
\affiliation{Diamond Light Source,
Didcot OX11 0DE, UK}

\author{Timur~K.~Kim}
\affiliation{Diamond Light Source,
Didcot OX11 0DE, UK}

\author{Cephise~Cacho}
\affiliation{Diamond Light Source,
Didcot OX11 0DE, UK}

\author{Federico~Mazzola}
\affiliation{CNR-IOM, Area Science Park, Strada Statale 14 km 163.5, I-34149 Trieste, Italy}

\author{Jun~Fujii}
\affiliation{CNR-IOM, Area Science Park, Strada Statale 14 km 163.5, I-34149 Trieste, Italy}

\author{Ivana~Vobornik}
\affiliation{CNR-IOM, Area Science Park, Strada Statale 14 km 163.5, I-34149 Trieste, Italy}

\author{Stuart~S.P.~Parkin}
\affiliation{Max Planck Institut f\"ur 
Mikrostrukturphysik, Weinberg 2, 06120 Halle, Germany}

\author{Barry~Bradlyn}
\affiliation{Department of Physics,
University of Illinois,
Urbana-Champaign, USA}

\author{Claudia~Felser}
\affiliation{Max Planck Institute for Chemical Physics of Solids,
Dresden
}
\author{Maia~G.~Vergniory}
\affiliation{Max Planck Institute for Chemical Physics of Solids,
Dresden
}
\affiliation{Donostia International Physics Center,
20018 Donostia - San Sebastian, Spain}

\author{Niels~B.~M.~Schr\"oter}
\email{niels.schroeter@mpi-halle.mpg.de}
\affiliation{Max Planck Institut f\"ur 
Mikrostrukturphysik, Weinberg 2, 06120 Halle, Germany}

\title {Supplementary Information: Parallel spin-momentum locking in a chiral topological semimetal}


\maketitle

\makeatletter
\renewcommand{\theequation}{S\arabic{equation}}
\renewcommand{\thefigure}{S\the\numexpr\value{figure}}
\renewcommand{\bibnumfmt}[1]{[S#1]}
\renewcommand{\citenumfont}[1]{S#1}

\tableofcontents

\section*{Out-of-plane spin-polarization}

\begin{figure*}[hbt]
  \includegraphics[width=0.99\linewidth]{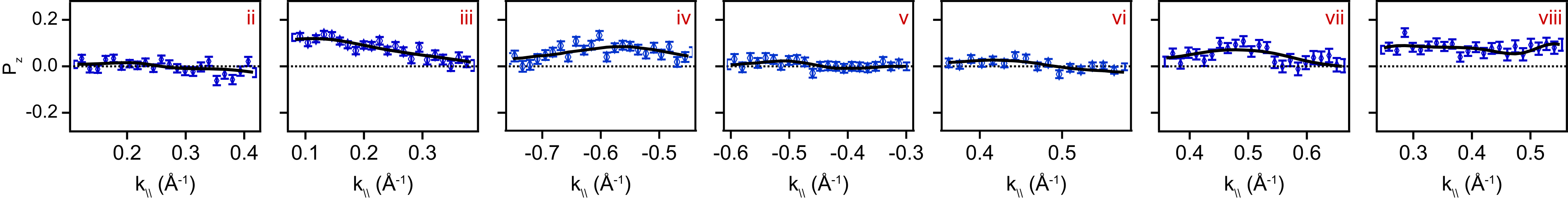}
  \caption{\textbf{Out-of-plane spin polarization}  Asymmetries in the out-of-plane spin-polarizations, corresponding to the cuts in Fig. 4(e) of the main text.
  }
\label{fig:Pz}
\end{figure*}

In contrast to the omnipresent in-plane spin-polarization, none of our spin-ARPES experiments showed any significant out-of-plane spin-polarization, as shown in Figure~\ref{fig:Pz}.

\begin{figure*}[hbt]
  \includegraphics[width=0.49\linewidth]{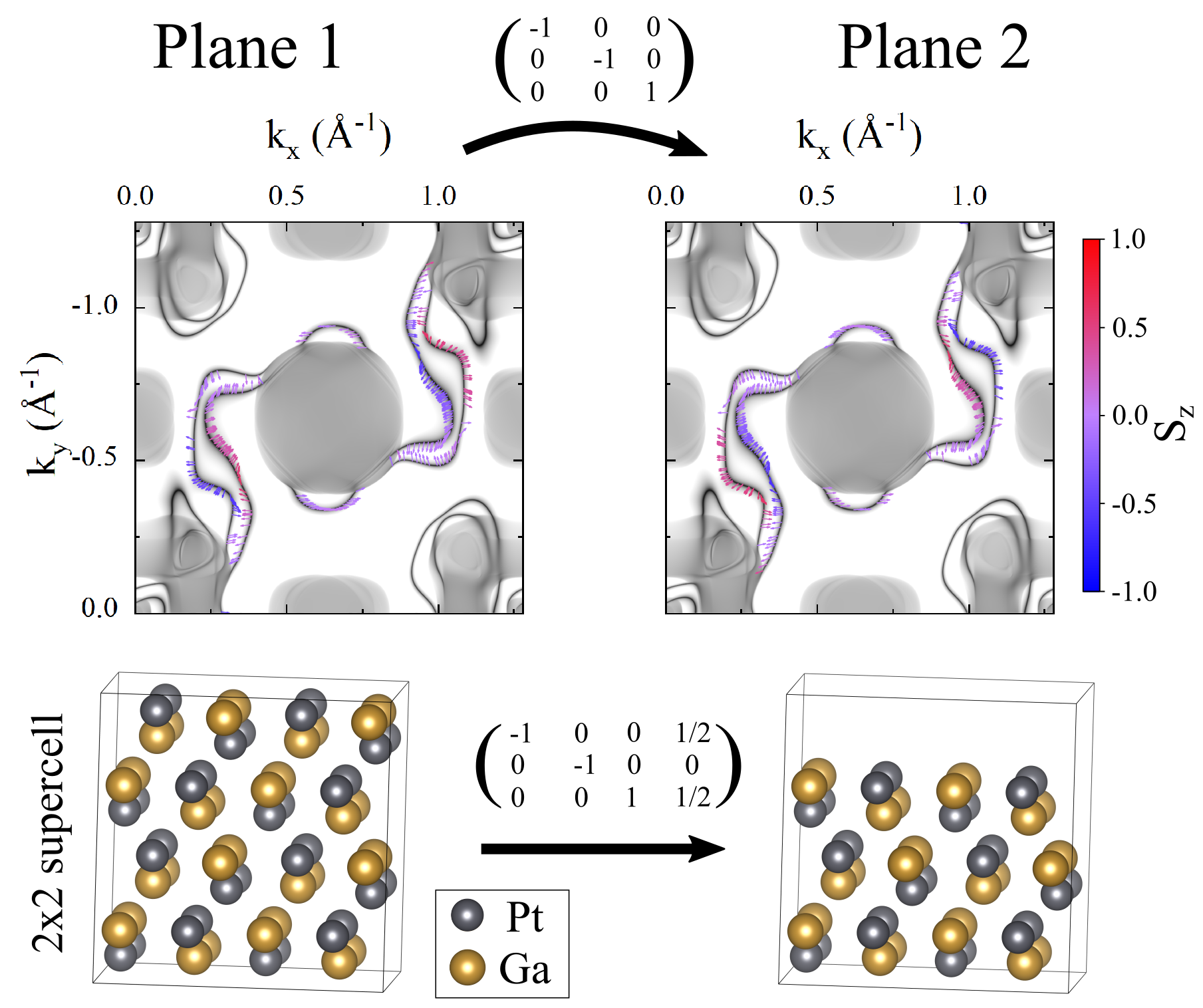}
  \caption{\textbf{Canceling out-of-plane surface spin polarization}  
  There are two different thermodynamically equivalent cleavage planes, related by a \SI{180}{\degree} roto-translation (bottom). The corresponding surface spin textures (top) have the same dispersion and in-plane spin texture, but opposite out-of-plane spin texture.
  }
\label{fig:cleaving}
\end{figure*}

This can be understood when considering the different terminations of PtGa(001). There are two thermodynamically equivalent  terminations, related by a \SI{180}{\degree} roto-translation as shown in Fig.~\ref{fig:cleaving} due to the screw symmetry axes along the principal directions in space group 198. We expect an approximately equal amount of domains from both planes to be present on our (001) polished crystal surfaces, as has been observed in the isostructural PdGa with STM~\cite{prinz_surface_2014}. While the in-plane spin texture is the same in both domains, the out-of-plane texture is opposite~(Fig.~\ref{fig:cleaving}). Since the ARPES beamspots were large compared to the expected domain size (tens of nm in PdGa~\cite{prinz_surface_2014}), the photoemmission signal will be the sum of the signal from both domains, and therefore have a vanishing out-of-plane spin polarization.

This can result in magnitudes of the measured Fermi-arc spin polarization smaller than 1. Note that the assumption of a polarization magnitude 1 for fitting the data has little effect on the direction of the resulting spin polarizations, but prevents the fit from being overparametrized.  

\section*{Spin-polarization in dependence of photon energy and polarization of light}

\begin{figure*}[hbt]
  \includegraphics[width=0.99\linewidth]{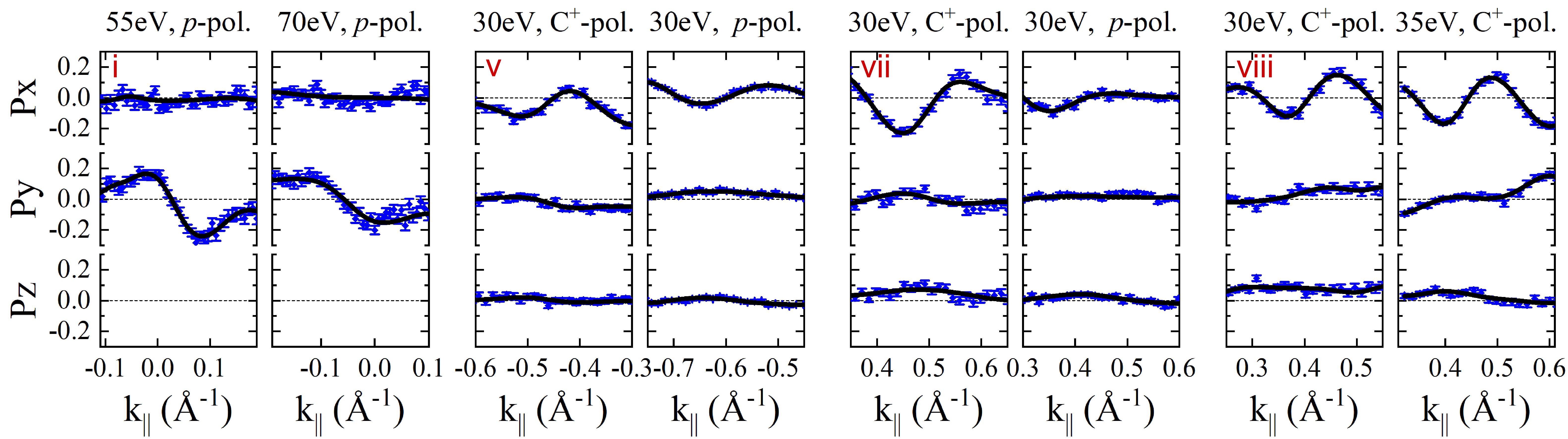}
  \caption{\textbf{Polarization and photon-energy dependence}  Selected spin cuts of Fig. 4(e) from the main text, measured with different photon energies and polarizations. 
  }
\label{fig:hv_pol}
\end{figure*}

To corroborate that the measured spin-texture is correctly reflecting the initial state spin polarization, we recorded several momentum distribution curves (MDCs) at different photon energies or light polarization, as shown in Figure~\ref{fig:hv_pol}. We find that the main spin polarization is always in the same channel, independent of the photon energy and light polarization used for the spin-ARPES experiments. We can therefore conclude that matrix element effects only play a minor role and that the measured spin-texture reflects the spin-texture of the initial state.

\section*{Control of spin-polarization \textit{via} PtGa enantiomorph}
\begin{figure*}[hbt]
  \includegraphics[width=0.42\linewidth]{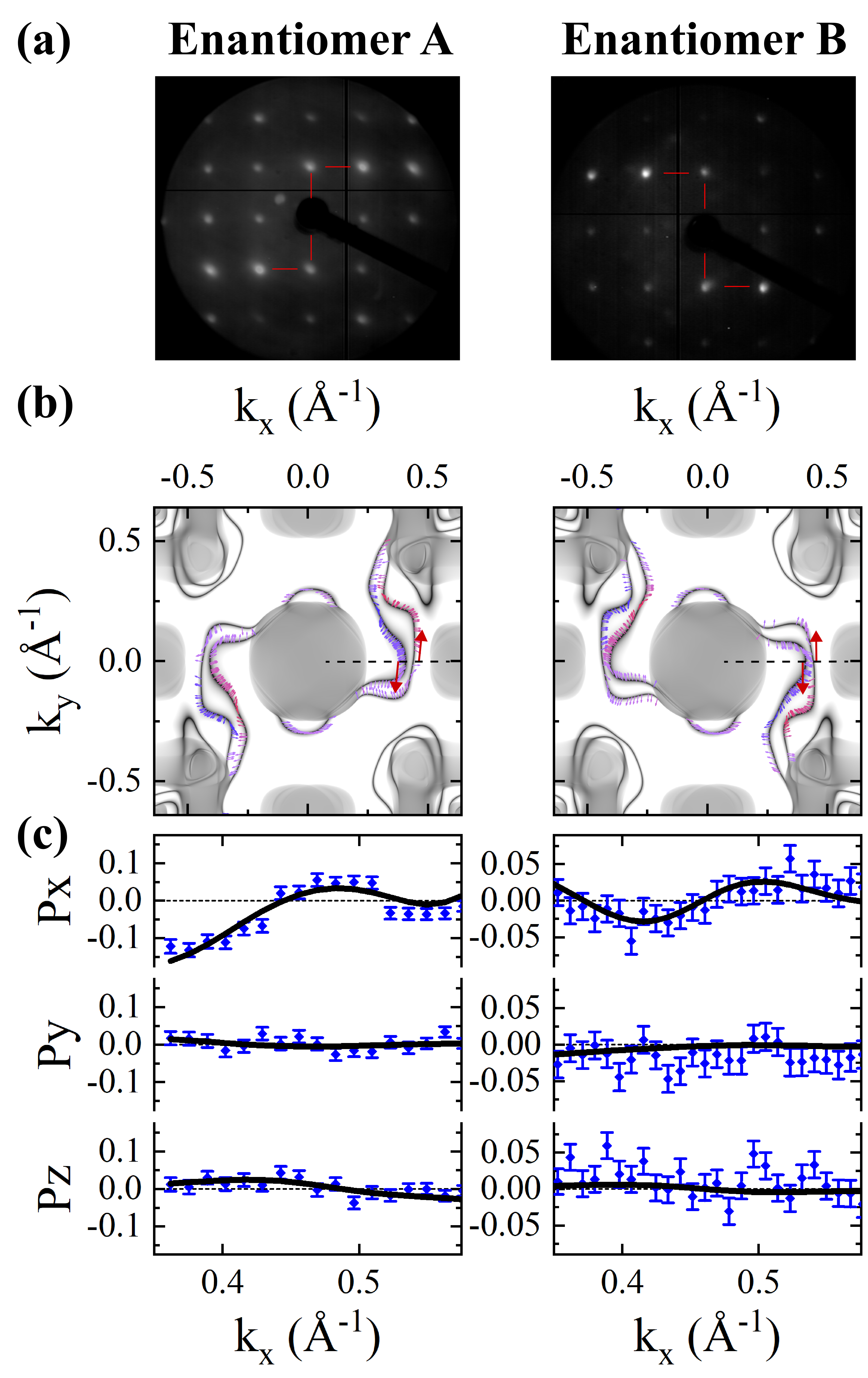}
  \caption{\textbf{Spin polarization in different enantiomers}  (a) LEED images at \SI{97.5}{eV} showing the different chiralities. (b) Calculated spin textures. (c)  Asymmetries in the measured spin-polarisations at \SI{30}{eV} with C+ polarisation in both enantiomers. The resulting spin directions are shown as red arrows in (b) The dispersions are mirrored, and the spin undergoes a pseudo-vector mirror. 
  }
\label{fig:enantiomorph}
\end{figure*}

The two enantiomorphs of PtGa are related by a mirror operation. The spin polarization behaves under such a mirror as a pseudo-vector, which we confirm by a measurement on the opposite enantomer, shown in Fig.~\ref{fig:enantiomorph}.

\section*{Band structure without spin-orbit coupling}
\begin{figure*}[hbt]
  \includegraphics[width=0.42\linewidth]{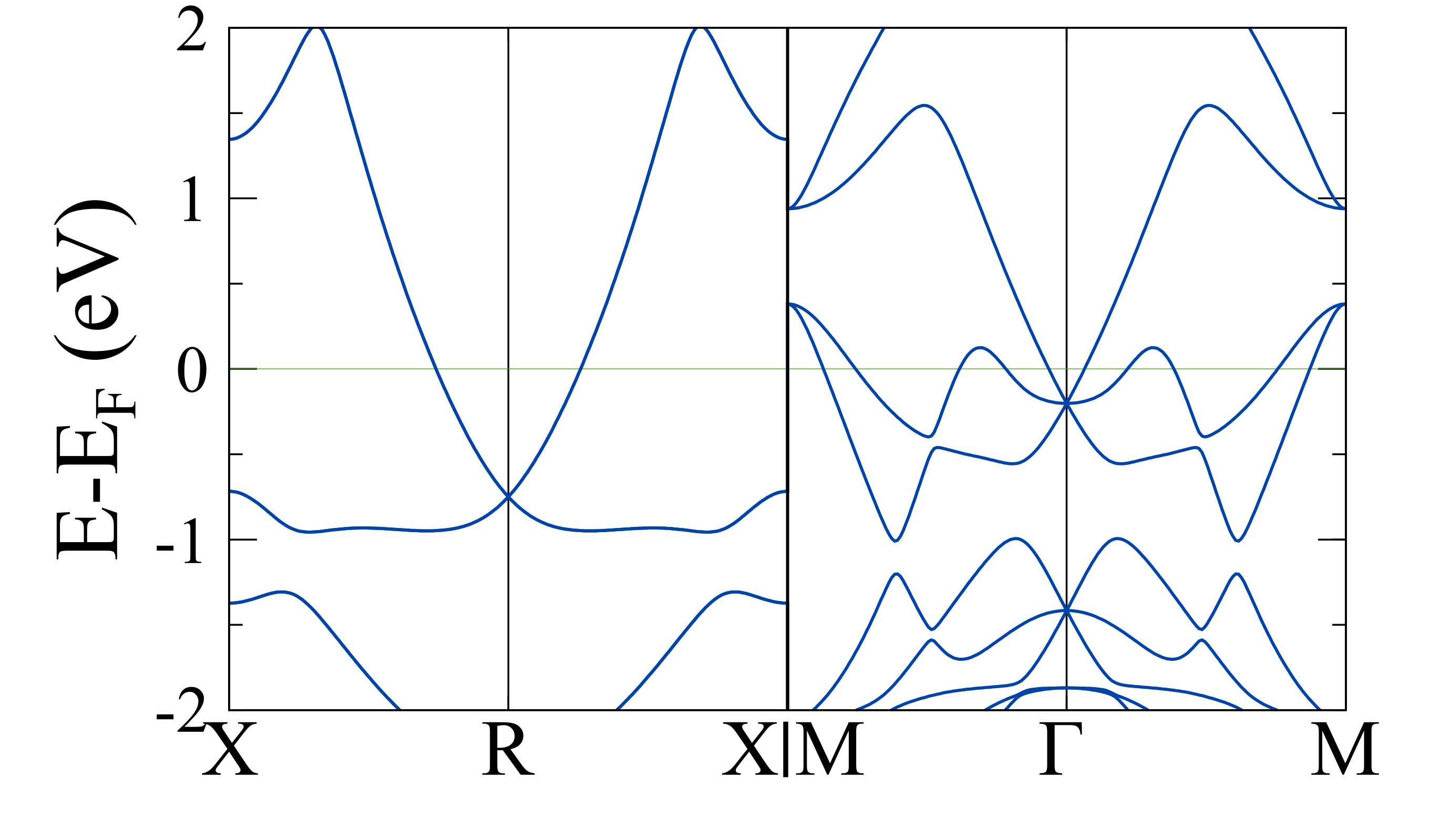}
  \caption{\textbf{Band structure without SOC}   Ab-initio calculated bandstructure without SOC along the same high symmetry directions shown in Fig.~2 of the main text.
  }
\label{fig:nsoc}
\end{figure*}

A reference band structure without SOC is shown in Fig.~\ref{fig:nsoc}

 \FloatBarrier
\bibliography{Biblio}